\documentclass[twocolumn]{emulateapj09}

\usepackage{graphicx}
\usepackage{epsfig}

\shorttitle{Lightcurves from Neutron Stars}
\shortauthors{Psaltis \& \"Ozel}

\begin{document}


\title{Pulse Profiles from Spinning Neutron Stars in the Hartle-Thorne
Approximation}

\author{Dimitrios Psaltis\altaffilmark{1} and Feryal \"Ozel\altaffilmark{1,2}}
\affil{Astronomy Department,
University of Arizona,
933 N.\ Cherry Ave.,
Tucson, AZ 85721, USA}
\email{dpsaltis, fozel@email.arizona.edu}

\altaffiltext{1}{also, Institute for Theory and Computation,
Harvard-Smithsonian Center for Astrophysics, 60 Garden St., Cambridge,
MA 02138}

\altaffiltext{2}{also, Radcliffe Institute for Advanced Study, Harvard
University, 8 Garden St., Cambridge, MA 02138}

\begin{abstract}
We present a new numerical algorithm for the calculation of pulse
profiles from spinning neutron stars in the Hartle-Thorne
approximation. Our approach allows us to formally take into account
the effects of Doppler shifts and aberration, of frame dragging, as
well as of the oblateness of the stellar surface and of its quadrupole
moment.  We confirm an earlier result that neglecting the oblateness
of the neutron-star surface leads to $\simeq 5-30$\% errors in the
calculated profiles and further show that neglecting the quadrupole
moment of its spacetime leads to $\simeq 1-5$\% errors at a spin
frequency of $\simeq 600$~Hz. We discuss the implications of our
results for the measurements of neutron-star masses and radii with
upcoming X-ray missions, such as NASA's NICER and ESA's LOFT. 
\end{abstract}

\keywords{stars: neutron --- relativity --- gravitation}

\section{INTRODUCTION}

A temperature anisotropy on the surface of a spinning neutron star
leads to a periodic oscillation of its brightness at the stellar spin
frequency, as viewed by an observer at infinity. The amplitude and
spectrum of this oscillation depends on the temperature profile on the
stellar surface, on the beaming of the emerging radiation, and on the
magnitude of strong-field gravitational lensing experienced by the
photons as they propagate through the neutron-star spacetime
(Pechenick, Ftaclas, \& Cohen 1983). Given a model for the emerging
radiation, the properties of the brightness oscillation can,
therefore, be used in mapping the neutron-star surface and spacetime,
as well as in measuring its mass and radius.

Pulse-profile modeling techniques have been used to explore the surface
emission properties in many types of neutron stars, such as slow
pulsars (Page 1995), magnetars (DeDeo, Psaltis, \& Narayan 2001;
\"Ozel, Psaltis, \& Kaspi 2001), rotation-powered millisecond pulsars
(Pavlov \& Zavlin 1997; Bogdanov, Rybicki, \& Grindlay 2007), X-ray
bursters (Weinberg, Miller, \& Lamb 2001; Nath, Strohmayer, \& Swank
2002; Muno, \"Ozel, \& Chakrabarty 2002, 2003), and accretion-powered
millisecond pulsars (Poutanen \& Gierlinski 2003; Bhattacharyya et
al.\ 2005; Leahy et al.\ 2008, 2009, 2001; Lamb et al.\ 2009; Morsink
\& Leahy 2011). This technique also defines the key scientific
objectives of several upcoming or proposed X-ray missions, 
such as NASA's NICER (Arzoumanian et al.\ 2010), ESA's LOFT 
(Ferroci et al.\ 2012), and ISRO’s Astrosat (Agrawal 2006), 
which aim to measure the masses and radii of several millisecond 
rotation-powered pulsars and X-ray bursters with high precision.

The effects of the gravitational lensing on the surface photons have
been explored with a variety of techniques and approximations, since
the original work on non spinning neutron stars (Pechenick, Ftaclas,
\& Cohen 1983). At relatively low spin frequencies (i.e., $\lesssim
300$~Hz), the spacetime of the neutron star is accurately described by
the Schwarzschild metric and its spin primarily causes Doppler shifts
and aberration (Miller \& Lamb 1998; Muno, \"Ozel, \& Chakrabarty
2002; Poutanen \& Beloborodov 2006). At similar spin frequencies, the
effects of frame dragging are marginal (Braje, Romani, \& Rauch 2000).
At moderate spin frequencies (i.e., $\lesssim 800$~Hz), the
neutron-star spacetime acquires a quadrupole moment and its surface
becomes oblate.  Finally, at spin frequencies near breakup (i.e.,
$\gtrsim 1$~kHZ), higher order multipoles become important and the
neutron-star spacetimes can only be calculated by solving numerically
the field equations for particular equations of state (Cook et al.\ 1994;
Stergioulas \& Friedman 1995; Stergioulas 2003; see also 
Cadeau et al.\ 2007).

Rotation powered millisecond pulsars and thermonuclear X-ray bursters, 
which make up the majority of neutron star targets for NICER and LOFT, 
have fast spin frequencies in the $\simeq 300-800$~Hz range. Stars 
with these spin frequencies, especially if they also possess larger radii, 
acquire notably oblate shapes and large quadrupole moments. 
Morsink et al.\ (2007) demonstrated that, for this relevant range of 
parameters, the oblateness of the neutron-star surface significantly 
affects the resulting pulse profiles. In a similar range of frequencies,
Baub\"ock, Psaltis, \& \"Ozel (2013) showed that the quadrupole moment
of a neutron-star spacetime also significantly alters the
spectroscopic properties of its surface emission.

In the present article, we describe a new algorithm for calculating
the lightcurves of spinning neutron stars in the Hartle-Thorne
approximation, which formally takes into account the oblateness and
quadrupole moment of the star, as well as the effects of frame
dragging, Doppler shift, and aberration. Contrary to calculations that
are based on numerical spacetimes, our approach allows us to simulate
lightcurves based only on the macroscopic properties of the neutron
stars, without the need for assuming a particular equation of state.

In \S2, we describe our calculations that are based on our ray-tracing
algorithm described in Psaltis \& Johannsen (2012) and Baub\"ock et
al.\ (2012). In \S3, we compare our calculations with earlier results
obtained under different assumptions and approximations. Finally, in 
\S4, we discuss the implications of our results for the mass and radius
measurements with the upcoming X-ray missions.

\section{Calculating Lightcurves of Spinning Neutron Star}

\begin{figure}[t]
\psfig{file=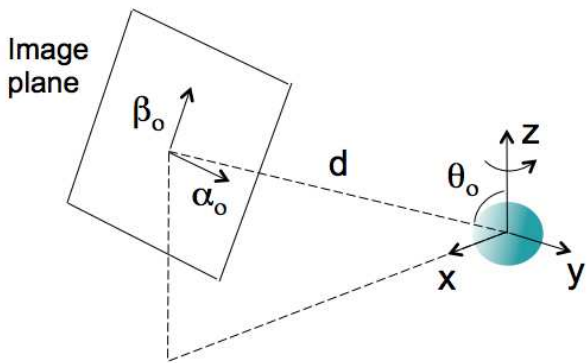,width=3.5in,clip=}
\caption{The geometry used in calculating the lightcurves from 
spinning neutron stars.}
\label{fig:geometry}
\end{figure}

Our goal is to calculate the brightness and the spectrum of emission
from the surface of a spinning neutron star at different rotational
phases and for different viewing geometries. To this end, we
define a coordinate system with its origin at the center of the
neutron star and the z-axis aligned with the rotation axis of the
star.  We then set up an image (or detector) plane at a distance $d$
from the center and at an angle $\theta_0$ with respect to the
rotational axis of the star.  At $\theta_0 = 0$, the image plane is
parallel to the $x-y$ plane, while for $\theta_0 = \pi/2$, the 
vector $d$ lies along the $x-$axis and the image plane is parallel to
the $y-z$ plane. On the image plane, we also define a two-dimensional
Cartesian coordinate system $(\alpha_0,\beta_0)$, with the
$\beta_0$-axis on the $x-z$ plane and the $\alpha_0$-axis pointing
towards the $y$-axis (see Figure~\ref{fig:geometry}).

The flux of radiation at photon energy $E$ received on the image plane
at time $t$ is given by the integral
\begin{equation}
F_E(t)=\frac{1}{d^2}\int_{\alpha_0,\beta_0} I(\alpha_0,\beta_0;E,t)
d\alpha_0d\beta_0\;,
\label{eq:flux}
\end{equation}
where $I(\alpha_0,\beta_0;E,t)$ is the specific intensity on the image
plane of a point with coordinates $(\alpha_0,\beta_0)$. Of all the
photon rays that reach the image plane, only those that originate on
the neutron-star surface have a non-zero intensity. Because of the
strong gravitational field of the neutron star, these photon rays are
not straight but are rather curved due to gravitational lensing.

The radii of all realistic neutron stars are larger than the radius of
the photon orbit in their spacetimes. Under these conditions,
photons arriving at position ($\alpha_0, \beta_0$) on the image plane
are uniquely connected to a position and direction on the stellar
surface. As a result, we can define a number of one-to-one maps
between the coordinates $(\alpha_0,\beta_0)$ at which a photon ray
crosses the image plane and various other properties of the photons
that travel along this ray. In particular, we will use the five maps
\begin{eqnarray}
\phi&=&\phi(\alpha_0,\beta_0)\label{eq:map1}\\
\theta&=&\theta(\alpha_0,\beta_0)\label{eq:map2}\\
\delta&=&\delta(\alpha_0,\beta_0)\label{eq:map3}\\
\epsilon&=&\frac{E(\alpha_0,\beta_0)}{E_{\rm e}(\phi,\theta)}
\label{eq:map4}\\
t_{\rm d}&=&t_{\rm d}(\alpha_0,\beta_0)\;,\label{eq:map5}
\end{eqnarray}
that connect the image-plane coordinates $(\alpha_0,\beta_0)$ of a
photon ray to the longitude $\phi$ and latitude $\theta$ on the
neutron star surface where the ray originates, to the angle $\delta$
between the photon momentum on the stellar surface and the normal to
the surface, to the ratio $\epsilon$ between the observed and emitted
energies, $E$ and $E_{\rm e}$, and to the time of flight between the
neutron star surface and the image plane.

Because of the Lorentz invariance of the photon occupation number
along a photon ray, we can now relate the specific intensity
$I(\alpha_0,\beta_0;E,t)$ on the image plane to the specific intensity
$I_{\rm NS}(\theta,\phi,\delta;E,t)$ on the neutron star surface as
\begin{eqnarray}
I(\alpha_0,\beta_0;E,t)&=&\epsilon^3 I_{\rm NS}
\left[\phi(\alpha_0,\beta_0),\theta(\alpha_0,\beta_0),\right.\nonumber\\
&&\qquad\left.
\delta(\alpha_0,\beta_0); \frac{E}{\epsilon}, 
t-t_{\rm d}(\alpha_0,\beta_0)\right]\;.
\end{eqnarray}
Inserting this last expression into equation~(\ref{eq:flux}) leads to
an integral expression for the time- and energy-dependent radiation
flux that flows through the image plane.

\subsection{Ray Tracing}

We calculate the five mapping
relations~(\ref{eq:map1})-(\ref{eq:map5}) using the ray tracing
algorithm described in Psaltis \& Johannsen (2012) and Baub\"ock et
al.\ (2012). In this algorithm, we describe the external spacetime of
a neutron star spinning at a moderate rate using the variant of the
Hartle-Thorne metric developed by Glampedakis \& Babak (2006).

The metric coefficients depend on the mass $M$ of the neutron star, on
its specific angular momentum $a$, and on the mass quadrupole moment
$q$ of the spacetime. We choose to write the quadrupole moment as
\begin{equation}
q=-a^2(1+\eta)\;,
\label{eq:quadrupole}
\end{equation}
for two reasons. First, when $\eta=0$, the quadrupole moment of the
spacetime reduces to that of the Kerr metric. Second, calculations of
the quadrupole moments of spinning neutron stars using numerical
algorithms (Laarakkers \& Poisson 1998; Pappas \& Apostolatos 2012)
showed that equation~(\ref{eq:quadrupole}) remains valid even for
neutron stars that are spinning near their breakup points. Typical
values of the quadrupole moment require $\eta\sim 1-6$, depending on
the equation of state, as well as on the stellar mass and radius.

\begin{figure}[t]
\psfig{file=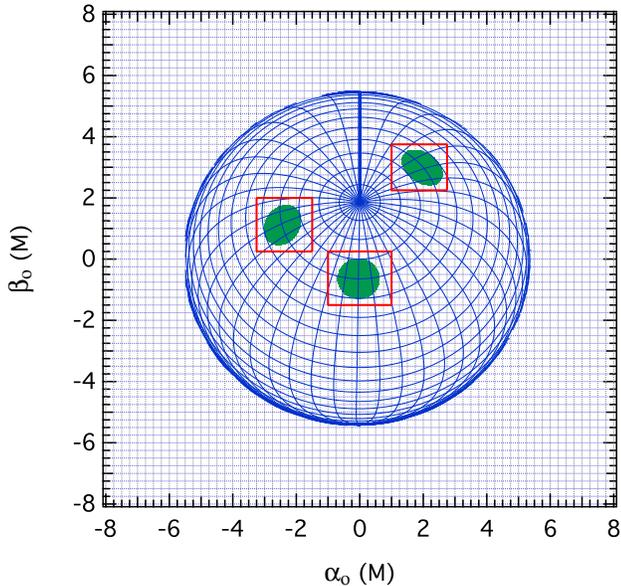,width=3.5in,clip=}
\caption{An illustration of the two grids used in a typical
  calculation of a lightcurve from a spinning neutron star. The image
  plane has a size of $16 GM/c^2$ on which we have set up a $65\times
  65$ grid. Contours of constant latitude and longitude are shown in
  $10^\circ$ intervals on the neutron star surface. A $10^\circ$ hot
  spot on the neutron star is shown at a colatitude of 40$^\circ$ and
  at three different rotational phases: $\omega t=0$, $\pi/2$, and
  $2\pi/3$. The red rectangles outline the regions in the
  low-resolution grid within which we set up a $256\times 256$
  high-resolution grid, for each rotational phase. In this figure,
  the inclination of the observer is 30$^\circ$, the neutron
  star has a mass of 1.8~$M_\odot$, a radius of 10~km, other
  parameters that are typical of the FPS equation of state, and is
  spinning at 600~Hz in the clockwise direction.}
\label{fig:grid}
\end{figure}

The shape of a neutron star spinning at moderate rates deviates from
spherical. In the Hartle-Thorne approximation, the dependence of the
neutron-star radius on the polar angle $\theta$ can be written in
terms of two parameters, which we choose to be the equatorial and the
polar radius of the star, $R_{\rm p}$ and $R_{\rm eq}$, as
\begin{equation}
\frac{R(\theta)}{R_{\rm eq}}=\sin^2\theta +
\frac{R_{\rm p}}{R_{\rm eq}}\cos^2\theta\;.
\end{equation}
For the purposes of the calculations reported in this paper, we use
the analytic fitting formula for the ratio of the polar to the
equatorial radius devised by Morsink et al.\ (2007) as discussed in
Baub\"ock et al.\ (2012). Moreover, the equatorial radii that we are
reporting are defined such that the proper circumference of a circle
in the equator at radius $R_{\rm eq}$ is equal to $2\pi R_{\rm eq}$. 
In all the simulations presented hereafter, the value of the neutron 
star radius quoted represents $R_{\rm eq}$ at the specified spin 
frequency. 

Finally, we need to specify the intensity of radiation on the stellar
surface. In this paper, we will assume that radiation emerges only
from a small circular spot of angular radius $\rho$ that is fixed at a
colatitude $\theta_{\rm s}$ with respect to the rotational pole of the
neutron star. The spectrum of the emerging radiation is that of a
blackbody of temperature $T_{\rm NS}$ and the emission is
isotropic. Because we will be reporting our results with the photon
energy expressed in units of $T_{\rm NS}$, the latter quantity will
not represent an additional parameter in our calculations.

In summary, the lightcurve from a spinning neutron star in the
Hartle-Thorne approximation and with the approximations discussed
above depends on the following parameters: {\em (i)\/} the mass $M$ of
the neutron star; {\em (ii)\/} the equatorial radius $R_{\rm eq}$ of
the neutron star; {\em (iii)\/} its specific angular momentum $a$;
{\em (iv)\/} the quadrupole moment of its spacetime as measured by the
parameter $\eta$; {\em (v)\/} the observer inclination $\theta_0$;
{\em (vi)\/} the colatitude of the spot $\theta_{\rm s}$; and {\em
  (vii)\/} the angular radius of the spot $\rho$.

\subsection{Discretization, Convergence, and Performance}

\begin{figure}[t]
\psfig{file=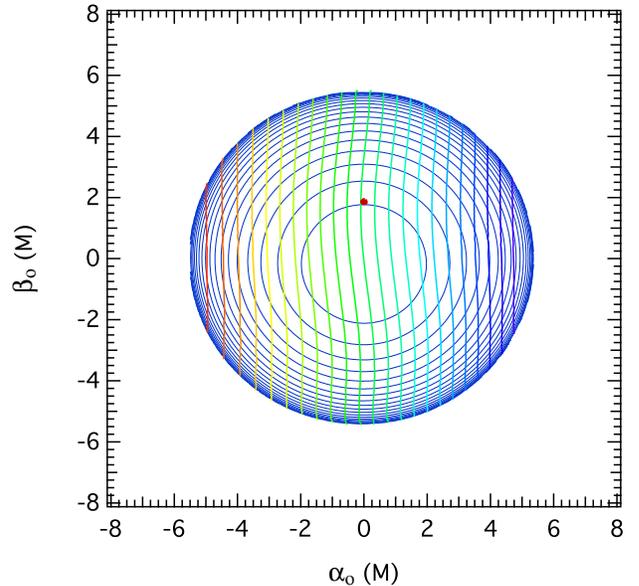,width=3.5in,clip=}
\caption{Contours of constant {\em (blue curves)\/} time delay and
  {\em (rainbow curves)\/} ratio of observed to emitted photon energy
  from the neutron star shown in Figure~\ref{fig:grid}. The time-delay
  contours are at equidistant intervals in the range $(0.5-10)\times
  GM/c^2$; the photon-energy contours are at equidistant intervals in
  the range $0.62-0.74$. The red dot marks the location of the north
  rotational pole of the neutron star.}
\label{fig:red_delay}
\end{figure}

In order to calculate efficiently the flux of radiation through the
image plane, especially when the size of the emitting region is very
small, we evaluate the integral~(\ref{eq:flux}) using a set of two
nested grids, as shown in Figure~\ref{fig:grid}. We initially use the
ray-tracing algorithm to calculate and store the five mapping
relations~(\ref{eq:map1})-(\ref{eq:map2}) on a low-resolution grid of
$N_l\times N_l$ points, with a typical value of $N_l=65$. For each
rotational phase of the neutron star, we use these results in order to
outline a rectangular region in the low-resolution grid that surrounds
the projection of the emitting region on the image plane (shown as red
rectangles in Figure~\ref{fig:grid} for three rotational phases). We
then set up a high-resolution grid within each of these rectangular
regions of $N_h\times N_h$ points, with a typical value of $N_l=256$.
Finally, we use the ray-tracing algorithm to calculate the mapping
relations on the high-resolution grid and evaluate numerically the
integral~(\ref{eq:flux}) only within this grid using a trapezoid
integration.

Attaining a sufficiently high resolution for the first three of
the five mapping relations~(\ref{eq:map1})-(\ref{eq:map5}) is what
drives the requirement for a high-resolution grid. This is shown in
Figure~\ref{fig:grid}, where contours of constant latitude and
longitude on the neutron star surface (in intervals of 10$^\circ$,
which is the same magnitude as the angular size of the spot in
this example) are projected on the image plane. The separation of
nearby contours decreases rapidly from the center of the stellar image
to its edge; therefore, the number of image plane points falling
within a spot near the edges also declines rapidly. The outline of
even a large spot of an angular radius of $10^\circ$, as in this
example, would be barely resolvable in the low-resolution grid shown
in the figure as the spot makes its way towards the edge of the
stellar image. In contrast, the last two mapping
relations~(\ref{eq:map4})-(\ref{eq:map5}) can be well approximated
even on the low-resolution grid. This is shown in
Figure~\ref{fig:red_delay}, in which contours of constant energy
ratios $\epsilon$ and time delays $t_{\rm d}$ are plotted on the image
plane. As expected, the contours of constant redshift are nearly
vertical curves spanning a narrow range, whereas the contours of
constant time delay are nearly concentric circles, centered at the
origin of the $\alpha_0,\beta_0$ plane. In more detail, the 
waviness of the redshift contours arises from two competing effects: 
the frame-dragging correction to the magnitude of the surface velocity 
as measured by a zero-angular-momentum observer and the quadrupole 
correction to the gravitational redshift. Note that the redshift 
contours are further distorted and develop the island and saddle 
shapes shown, e.g., in the right panel of Figure 5 of Baub\"ock et 
al.\ (2013), when the neutron star is viewed from a smaller inclination 
angle and has a larger quadrupole moment. The frame-dragging and the 
quadrupole corrections to the time-delay contours are typically 
less than $1/1000$ of the neutron star spin period and are, therefore, 
negligible. 

\begin{figure}[t]
\psfig{file=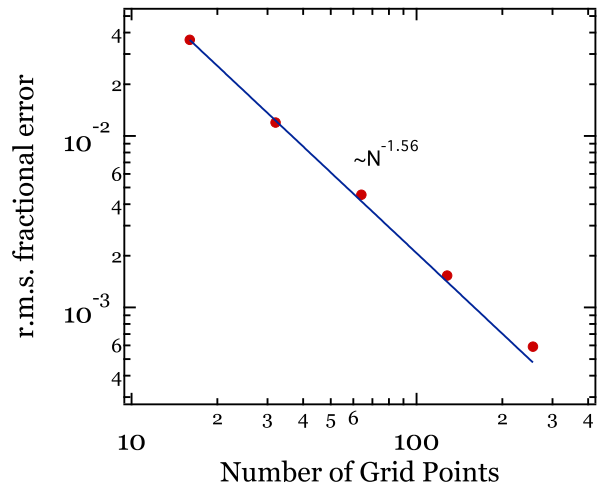,width=3.5in,clip=}
\caption{Convergence plot for the configuration shown in
  Figure~\ref{fig:grid}. Calculations with different numbers of points
  in the high-resolution grid are compared to a fiducial calculation
  that has $512\times 512$ grid points. The rms fractional error
  between their lightcurves in 32 phase bins is shown as a function of
  the number of grid points. The solid line shows the best fit slope
  of $\sim N^{-1.56}$, which is determined mainly by the ability of
  the 2D rectangular grid to trace the shape of the deformed circular
  spot.}
\label{fig:conv}
\end{figure}

The convergence of the ray-tracing algorithm was demonstrated in
Psaltis \& Johannsen (2012), where the results were also compared to
other algorithms for calculating the profiles of relativistically
broadened fluorescent lines around black holes. In
Figure~\ref{fig:conv}, we derive numerically the convergence rate of
our integration algorithm for the calculation of the radiation
flux. We have taken the calculation shown in Figure~\ref{fig:grid},
with a high-resolution grid of $512\times 512$ points nested in a
low resolution grid of $65 \times 65$ points as the fiducial one and
calculated the time-dependent flux at a photon energy of $E=3k_{\rm
B}T_{\rm NS}$ at 32 phase bins. We then repeated the same calculation
with $N_h = $16, 32, 64, 128, and 256 grid points along each dimension
of the high-resolution grid and compared these lower-resolution runs
with the fiducial calculation. We plot in Figure~\ref{fig:conv} the
r.m.s.\ fractional difference between each of the lower-resolution
runs and the fiducial run, as a function of the number of grid points
used. The blue line shows the best-fit power-law relation between the
two quantities plotted, with the fractional difference scaling
approximately as $\sim N_{\rm h}^{-1.56}$. This convergence rate is
determined predominantly by the ability of the two-dimensional grid to
trace the shape of the deformed circular spot, as the latter is
projected onto the image plane.

Figure~\ref{fig:conv} demonstrates that a high-resolution grid of
$128\times 128$ points leads to an accuracy of one part in thousand,
which is adequate for most applications. A typical calculation takes
approximately one second per phase bin on a fast workstation. This
time can be reduced by up to two orders of magnitude if the
ray-tracing part of the algorithm is performed on a GPU card (see
Chan, Psaltis, \& \"Ozel 2013).

\subsection{Comparison with Earlier Calculations}

\begin{figure}[t]
\psfig{file=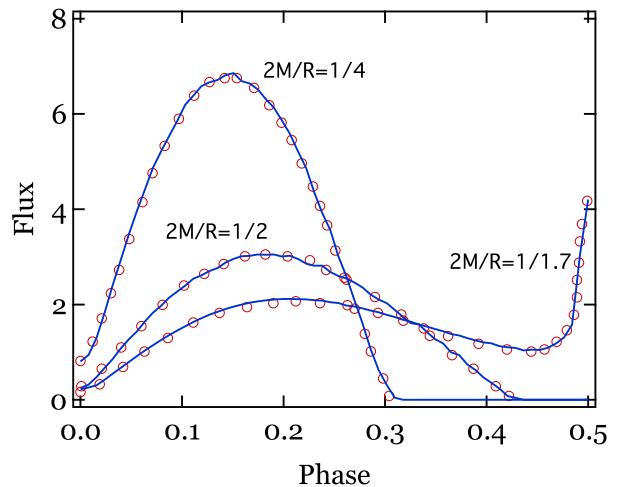,width=3.5in}
\caption{The flux of radiation from slowly-spinning neutron stars with
  different values of the compactness $2GM/Rc^2$, emitting from a
  single hot spot, as a function of the spin phase. The open circles
  are taken from Figure~10 of Pechenick et al.\ (1983), while the
  solid curves are the profiles calculated in the present work. The
  inclination of the observer is 90$^\circ$, the hot spot is on the
  stellar equator, its angular half-size is $5^\circ$, and the
  radiation emerges from its surface with a beaming proportional to
  $\sin\delta$.}
\label{fig:pechenick}
\end{figure}

Figure~\ref{fig:pechenick} compares three lightcurves for a slowly
spinning (1~Hz) neutron star calculated with the current algorithm to
the results presented by Pechenick et al.\ (1983) in their Figure~10.
For this particular comparison, the beaming of the emission that
emerges from the stellar surface was taken to be proportional to
$\sin\delta$.  The three different lightcurves correspond to neutron
stars with increasing compactness, from $2GM/Rc^2=1/4$ to
$2GM/Rc^2=1/1.7$. As expected, at this low spin frequency, the Doppler
effects, as well as those of frame dragging, oblateness, and of the
spacetime quadrupole moment are negligible.

\begin{figure}[t]
\psfig{file=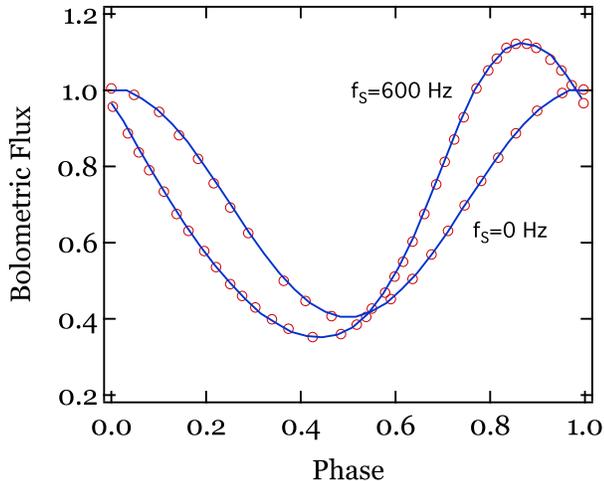,width=3.5in}
\caption{The bolometric flux of radiation from neutron stars with
  different spin frequencies, emitting from a single hot spot, in the
  Schwarzschild+Doppler approximation. The open circles are taken from
  Figure~2 of Poutanen \& Beloborodov (2003), while the solid curves
  are the profiles calculated in the present work. The inclination of
  the observer is 45$^\circ$, the colatitude of the hot spot is
  45$^\circ$, its angular half-size is $5^\circ$, and the radiation
  emerges from its surface with isotropic beaming. The neutron stars
  have a mass of 1.4~$M_\odot$ and a compactness of $2GM/Rc^2=1/2.5$.}
\label{fig:poutanen}
\end{figure}

\begin{figure*}[t]
\psfig{file=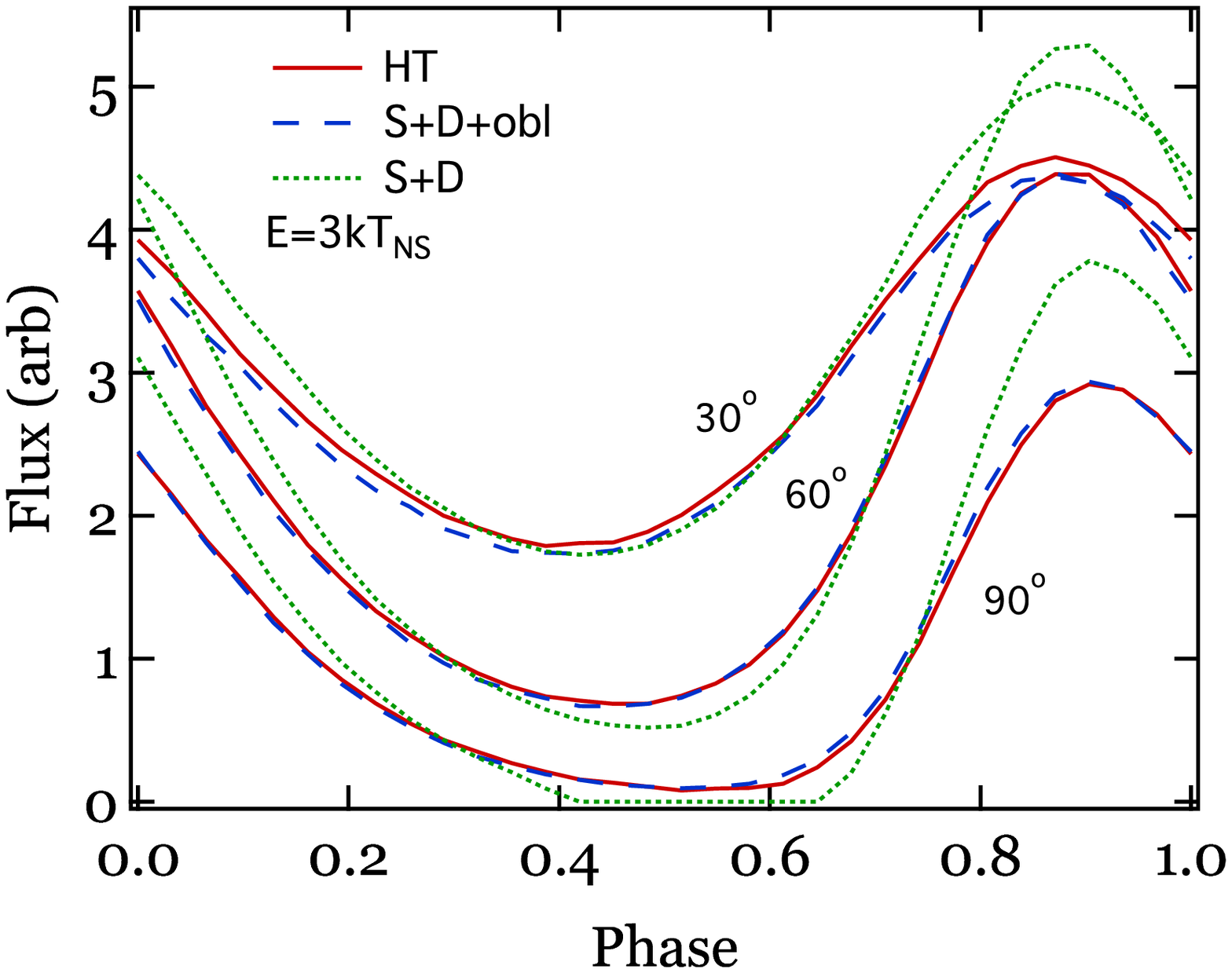,width=3.5in}
\psfig{file=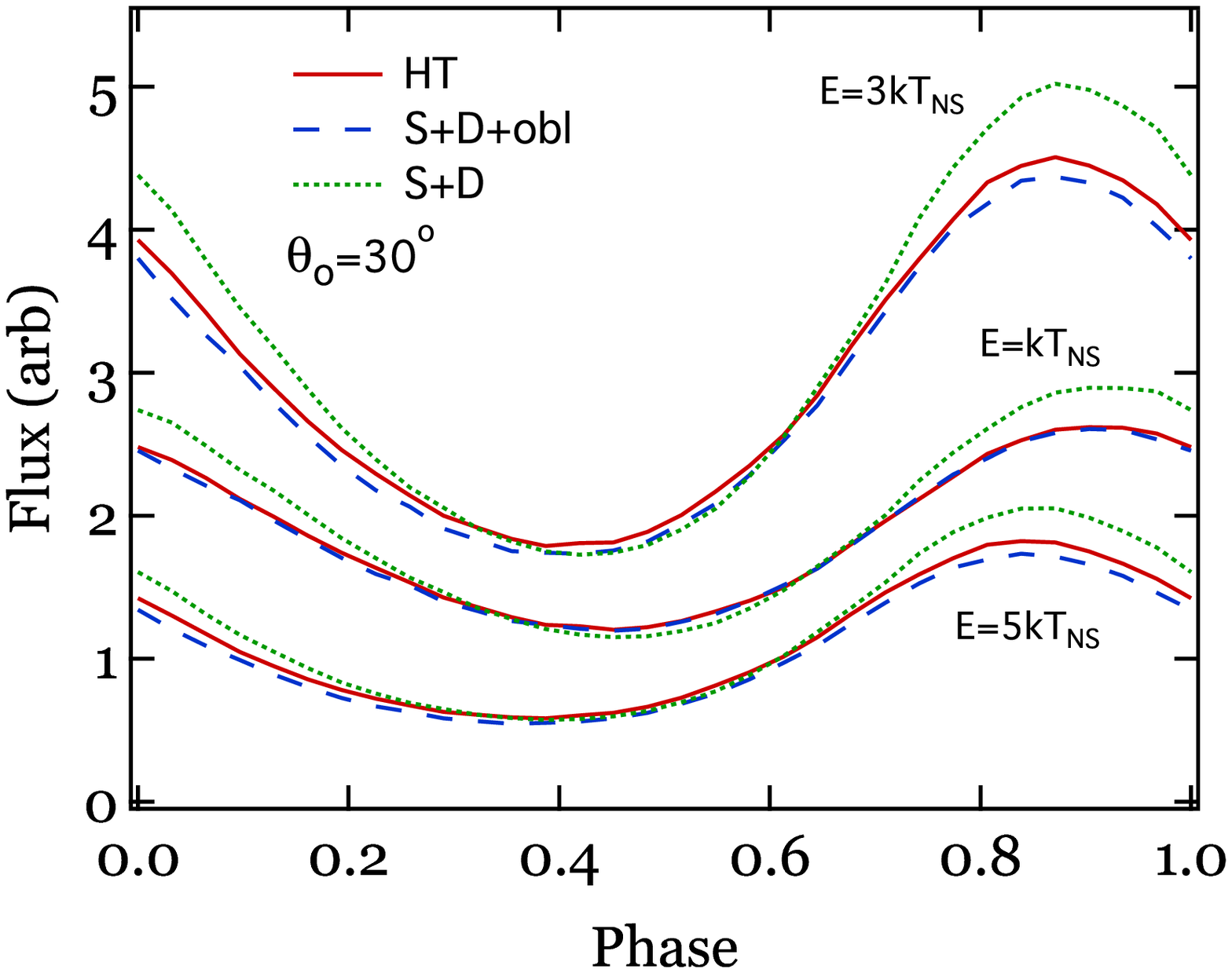,width=3.5in}
\caption{The flux of radiation received from a spinning neutron star
  {\em (Left)\/} for a photon energy equal to three times the
  neutron-star temperature but observed at different inclinations and
  {\em (Right)\/} at an observer inclination of $30^\circ$ but for
  three different photon energies. In both panels, the solid red line
  shows the calculation with the Hartle-Thorne metric, the dashed blue
  line shows the calculation in the Schwarzshild+Doppler approximation
  with the oblateness of the neutron star taken into account, and the
  dotted green line shows the calculation in the Scharzschild+Doppler
  approximation. All calculations correspond to a 1.8~$M_\odot$
  neutron star, with an equatorial radius $R_{\rm eq}$ fixed to
    15~km, spinning at 600~Hz, and with values for the remaining
  parameters that are typical for the L equation of state. The
  emitting region is taken to be a circular hot spot with a
  semi-angular size of $10^\circ$ and positioned at a colatitude of
  $40^\circ$ from the stellar rotational pole. The emission is that of
  a blackbody, with isotropic beaming. All fluxes have been multiplied
  by the same, constant factor for clarity.}
\label{fig:L600}
\end{figure*}

\begin{figure}[t]
\psfig{file=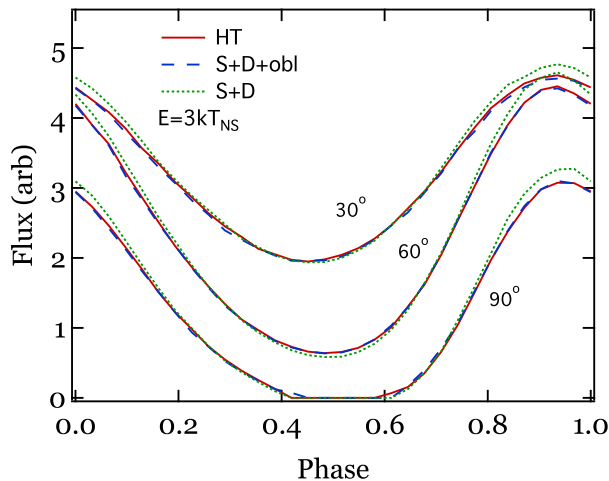,width=3.5in}
\caption{Same as in the left panel of Figure~\ref{fig:L600} but for a
neutron star spinning at 300~Hz.}
\label{fig:L300}
\end{figure}

\begin{figure}[t]
\psfig{file=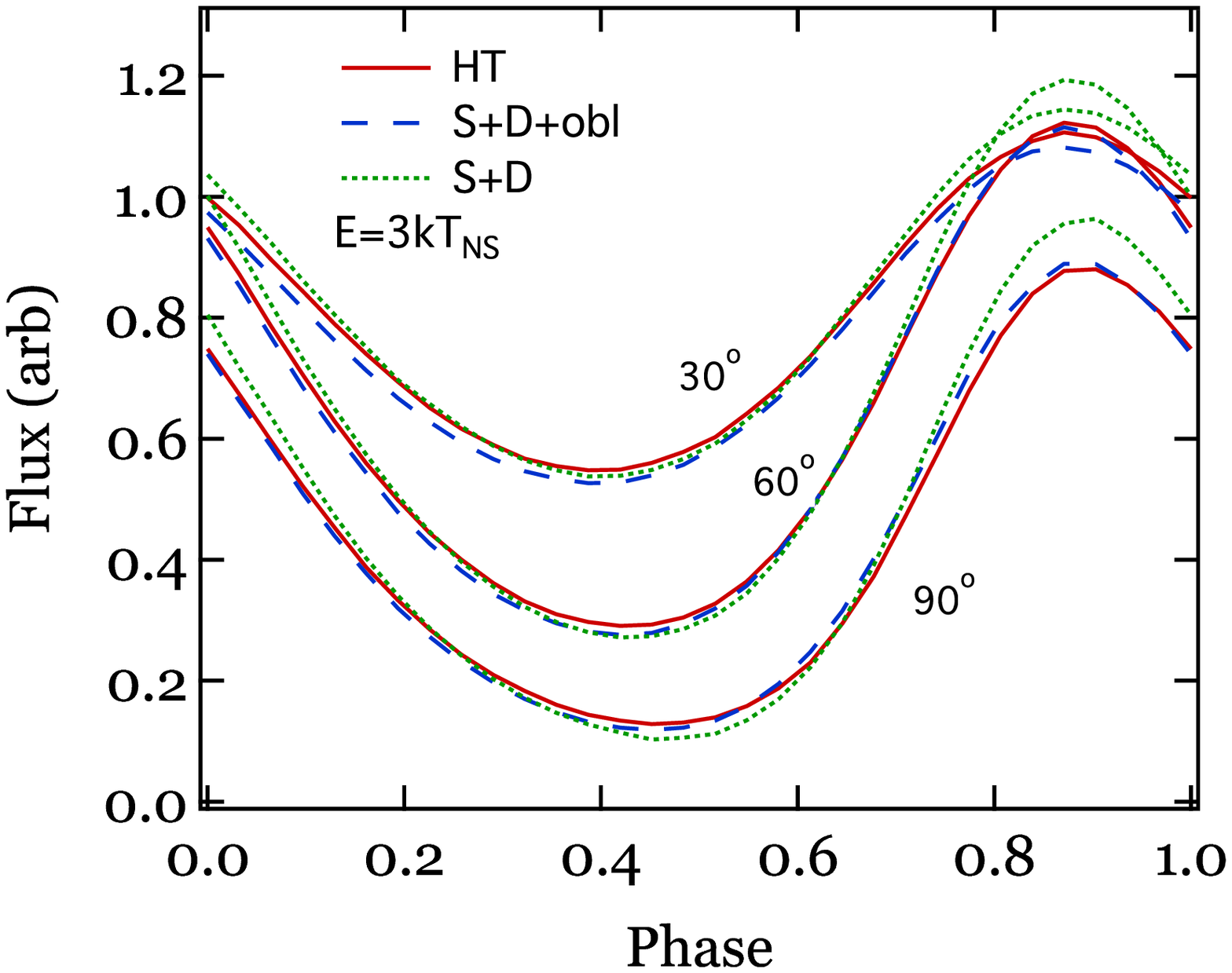,width=3.5in}
\caption{Same as in the left panel of Figure~\ref{fig:L600} but for a
  1.8~$M_\odot$, 10~km neutron star, spinning at 600~Hz, with
  parameters that are typical for the FPS equation of state.}
\label{fig:A600}
\end{figure}

Figure~\ref{fig:poutanen} compares two additional lightcurves with the
results shown in Figure~2 of Poutanen \& Beloborodov (2003), who used
the Schwarzschild+Doppler approximation. Because we consider this to
be a verification comparison, we artificially set the frame dragging,
the stellar oblateness, and the spacetime quadrupole to zero in our
calculations. As discussed in, e.g., Braje et al.\ (2000) and Poutanen
\& Boloborodov (2003), Doppler shifts and aberration have three main
effects on the pulse profiles: they increase their amplitudes, they
shift the location of their maxima to earlier phases, and they
introduce an asymmetry to the profile. Even though in our approach the
effects of Doppler shifts and of aberration are calculated
automatically and cannot be separated from those of the gravitational
lensing and of the gravitational redshift, our results agree, as
expected, with the earlier approximate methods.

\mbox{}

\section{Results}

\begin{figure*}[t]
\psfig{file=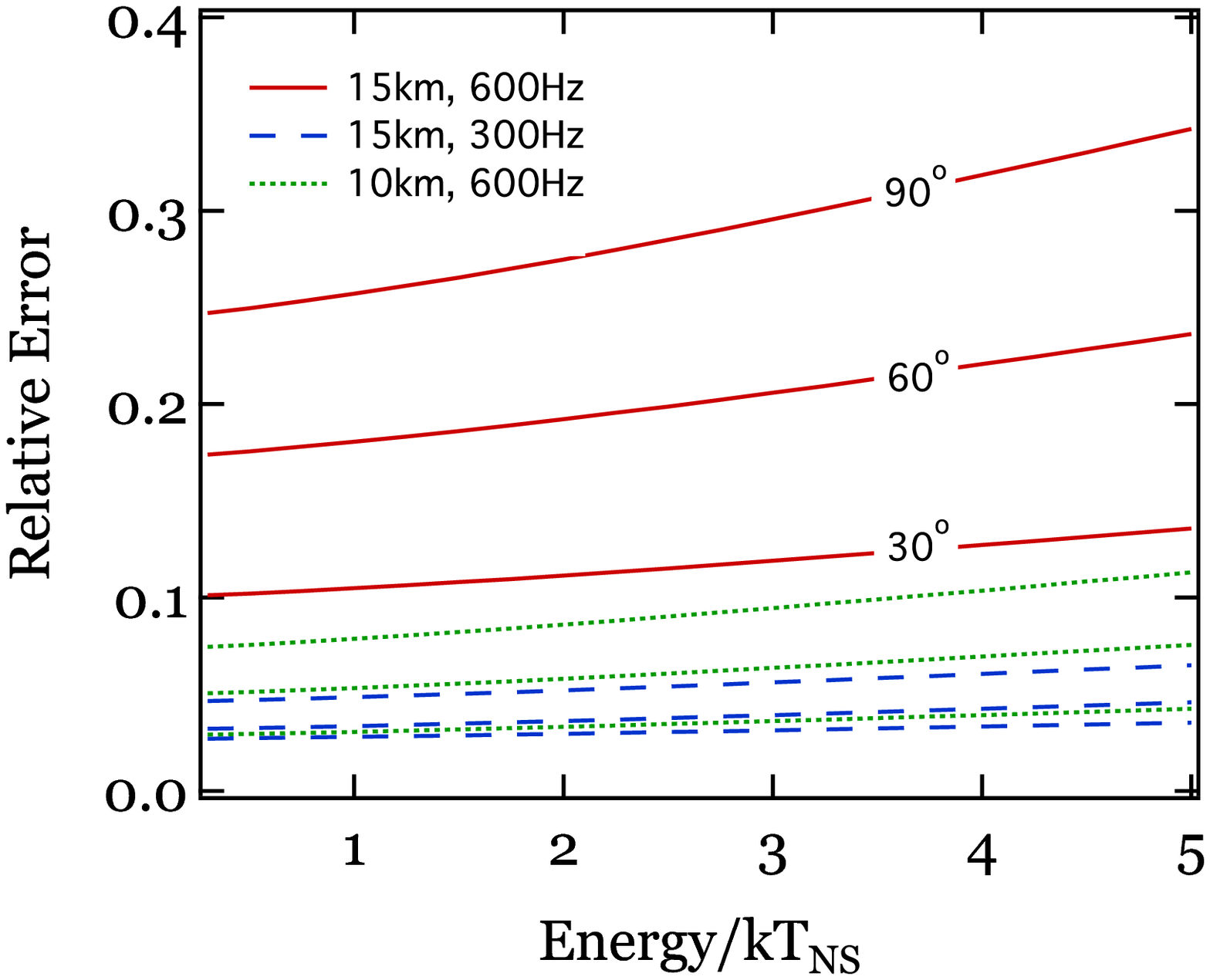,width=3.5in}
\psfig{file=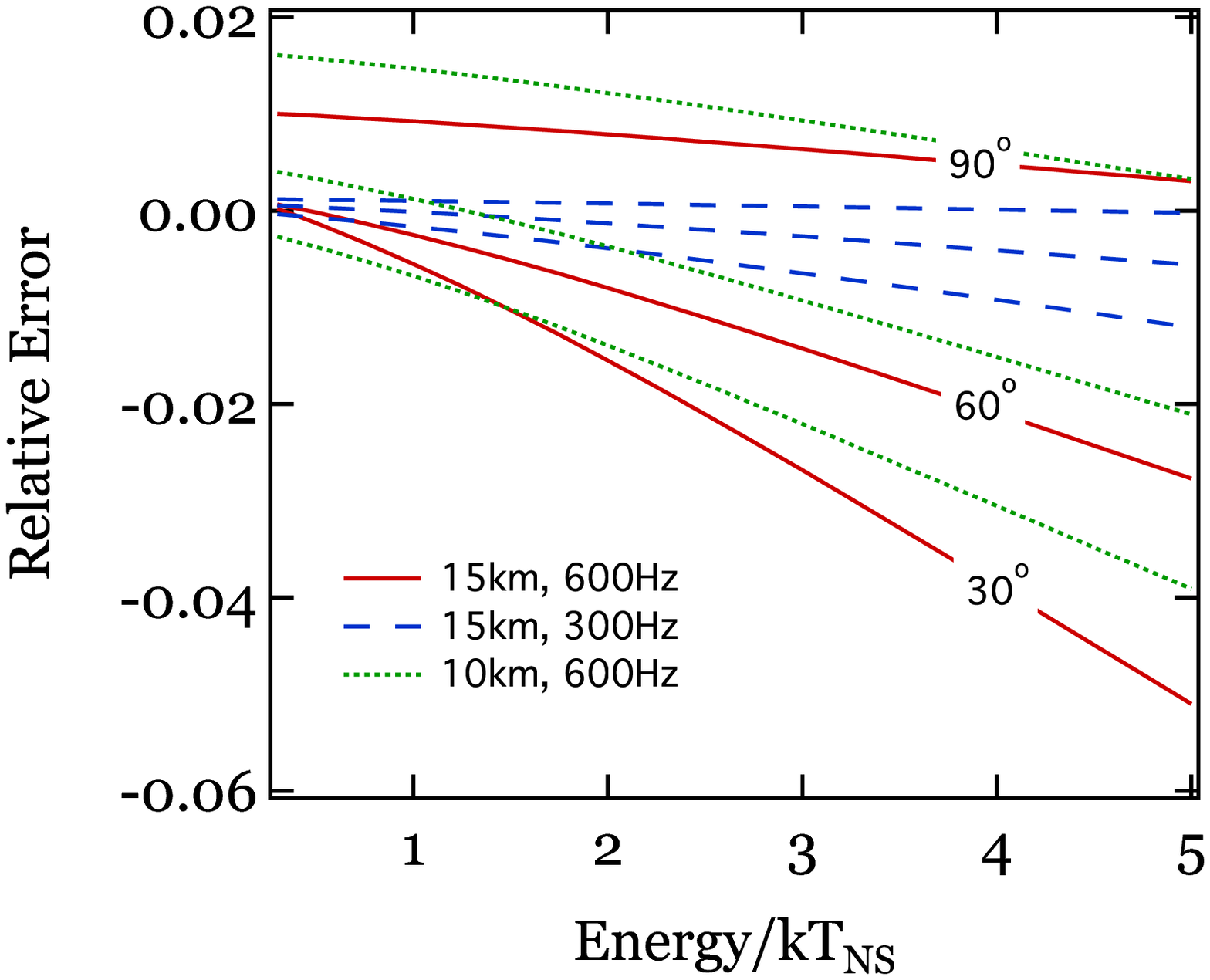,width=3.5in}
\caption{The energy dependence of the relative errors of the flux at
  spin phase 0.9 with the respect to the calculation in the
  Hartle-Thorne metric of {\em (Left)\/} the Schwarschild+Doppler
  approximation with spherical neutron stars and {\em (Right)\/} of
  the Schwarzschild+Doppler approximation with the oblateness of the
  neutron stars taken into account .  The various line styles
  correspond to the different calculations shown in
  Figures~\ref{fig:L600}-\ref{fig:A600}: the solid red curves are for
  the 15~km neutron stars spinning at 600~Hz shown in
  Figure~\ref{fig:L600}; the dashed blue curves are for the 15~km
  neutron star spinning at 300~Hz shown in Figure~\ref{fig:L300}; the
  dotted green curves for the 10~km neutron star spinning at 600~Hz
  shown in Figure~\ref{fig:A600}. In each case, the three curves of
  the same style correspond, from top to bottom, to observer
  inclinations of 90$^\circ$, 60$^\circ$, and 30$^\circ$,
  respectively.}
\label{fig:errors}
\end{figure*}

We now explore the effects of the stellar oblateness and of the
spacetime quadrupole moment on the pulse profiles.
Figure~\ref{fig:L600} shows the result of incorporating these two
effects, one at a time, on the pulse profile calculated for a
$1.8~M_\odot$, 15~km neutron star described by the equation of state L
and spinning at 600~Hz (see Baub\"ock et al. 2012 for a description of
the equations of state and their relevant parameters). This rather
large radius for the neutron star, albeit disfavored by current
observations (see \"Ozel 2013 and references therein), leads to the
largest Doppler, oblateness, and quadrupole effects. The left panel of
the figure shows the pulse profiles at three different observer
inclinations at a photon energy equal to three times the surface
temperature, which roughly corresponds to the peak of the spectrum
(before applying the gravitational and Doppler shifts). The right panel
shows the pulse profile at a single observer inclination but for three
different photon energies.

As discussed in Morsink et al.\ (2007), taking into account the
oblateness of the stellar surface significantly reduces the amplitude
of pulsations. Comparing the pulse profiles of two neutron stars with
the same equatorial radius but different surface shapes, an increasing
oblateness both increases the gravitational redshift experienced by
photons leaving a non-equatorial emitting region as well as alters the
relative orientation of the emitting region with respect to the
distant observer. The net effect is a decrease in the observed flux at
the peak of the pulse profile. Moreover, the oblate shape of the
stellar surface alters the visibility of different regions of the
surface by a distant observer and changes the flux at the minimum of
the pulse profile. The phase of the maximum flux and the asymmetry of
the profile, both of which are predominantly determined by Doppler
effects are not altered significantly. Taking into account the
oblateness of the stellar surface while still neglecting the
quadrupole of the spacetime corrects for most of the difference
between the Schwarzschild+Doppler and the Hartle-Thorne
approximations.

The quadrupole of the spacetime alters the pulse profiles primarily
through its effects on the gravitational redshift experienced by
photons. As discussed in Baub\"ock et al. (2013), the spacetime
quadrupole competes with the Doppler shifts in determining the energy
and brightness of the emerging radiation. As the observer inclination
is reduced, the Doppler effects become less important and the redshift
experienced by the photons is dominated by the spacetime quadrupole.

Figures~\ref{fig:L300} and ~\ref{fig:A600} show that the pulse
profiles for the same set up and observer inclinations, but for a
neutron star spinning at 300~Hz and for a 10~km neutron star,
respectively. These figures show that, as expected, the effects of the
stellar oblateness and of the spacetime quadrupole become less
important as the spin frequency of the neutron star or its equatorial
radius are reduced. The correction introduced by the oblateness of the
neutron star, albeit small, still dominates the correction introduced
by the spacetime quadrupole.

Figure~\ref{fig:errors} quantifies the magnitudes of the errors
introduced in the calculation of the pulse profiles by neglecting the
stellar oblateness and the spacetime quadrupole. As expected from the
above discussion, the effects of the stellar oblateness become large
as the observer inclination becomes very different from the colatitude
of the emitting region but have a weak dependence on photon energy.
On the other hand, the effects of the spacetime quadrupole have a
stronger energy dependence and become more significant (in absolute
value) as the inclination of the observer is reduced. This is
consistent with the fact that quadrupole effects arise primarily from
the influence of the spacetime quadrupole on the gravitational
redshift experienced by the photons and their relative contribution to
the observed flux increases as the Doppler effects become less
pronounced.

The overall effects of the stellar oblateness and of the spacetime
quadrupole for a 600~Hz spin frequency are of order $\sim 10-30$\% and
$\sim 1-5$\%, respectively, depending on the radius of the neutron
star and the observer inclination. These are comparable to the stated
5\% target uncertainty in the measurement of neutron-star masses and
radii using observations of pulse profiles with NICER and LOFT. As a
consequence, achieving the goals of these two missions requires
calculating pulse profiles with both these effects taken into account.

\acknowledgements

We thank Michi Baub\"ock for many useful discussions on ray tracing in
neutron-star spacetimes and the referee Sharon Morsink for enlightening 
discussions and comments.  This work was supported in part by NSF grant
AST-1108753, NSF CAREER award AST-0746549, and Chandra Theory grant
TM2-13002X. F.\"O. gratefully acknowledges support from the Radcliffe
Institute for Advanced Study at Harvard University.

\end{document}